\documentclass[journal=jacsat,manuscript=article] {achemso}
\usepackage[version=3]{mhchem} % Formula subscripts using \ce{}
\usepackage{graphicx}
\usepackage{float}
\usepackage[section]{placeins}
\usepackage[latin1]{inputenc}
\usepackage{amsmath}
\usepackage{amsfonts}
\usepackage{amssymb}
\usepackage{subfigure}
\usepackage{dcolumn}   % needed for some tables
\usepackage{bm}            % for math
\usepackage{graphicx}
\usepackage{float}
\usepackage[section]{placeins}
\usepackage{subfig}
\usepackage{color}

\newcommand{\onlinecite}[1]{\hspace{-1 ex} \nocite{#1}\citenum{#1}} 

\author{Marco Nava}
\affiliation
{Department of Chemistry and Applied Biosciences, ETH Zurich, and Facolt\`a di Informatica, Istituto di Scienze Computazionali, Universit\`a della Svizzera Italiana, Via G. Buffi 13, 6900 Lugano Switzerland}
\email{mark.nava@gmail.com}

\author{Ferruccio Palazzesi}
\affiliation
{Department of Chemistry and Applied Biosciences, ETH Zurich, and Facolt\`a di Informatica, Istituto di Scienze Computazionali, Universit\`a della Svizzera Italiana, Via G. Buffi 13, 6900 Lugano Switzerland}

\author{Claudio Perego}
\affiliation
{Department of Chemistry and Applied Biosciences, ETH Zurich, and Facolt\`a di Informatica, Istituto di Scienze Computazionali, Universit\`a della Svizzera Italiana, Via G. Buffi 13, 6900 Lugano Switzerland}

\author{Michele Parrinello}
\affiliation
{Department of Chemistry and Applied Biosciences, ETH Zurich, and Facolt\`a di Informatica, Istituto di Scienze Computazionali, Universit\`a della Svizzera Italiana, Via G. Buffi 13, 6900 Lugano Switzerland}
\email{parrinello@phys.chem.ethz.ch}

\title{Dimer Metadynamics}

\date{\today}

\begin{document}
\begin{abstract}
Sampling complex potential energies is one of the most pressing challenges of
contemporary computational science. Inspired by recent efforts that use 
quantum effects and discretized Feynman's path integrals to overcome large barriers we 
propose a replica exchange method. In each replica two copies of the same 
system with halved potential strengths interact via inelastic springs. The strength of the spring is varied in 
the different replicas so as to bridge the gap between the infinitely strong spring, that corresponds to  
the  Boltzmann replica  and the less  tight ones. We enhance the spring length 
fluctuations using Metadynamics. We test the method on simple yet challenging problems.
\end{abstract}
\maketitle

\section{Introduction}
The problem of sampling complex free energy landscapes that exhibit long lived metastable states separated by large barriers is of great current interest\cite{longscale1,longscale2,longscale4}. 
The vast literature on the subject is a clear evidence of its pressing relevance\cite{meta_review}.
Roughly speaking, two classes of methods can be identified. In one, a set of collective variables (CVs) that depend on the microscopic coordinates of the system 
is chosen, and a bias that depends on these chosen collective variables is constructed so as to speed up sampling. Examples of this approach are 
Umbrella Sampling\cite{met1}, Local Elevation\cite{met2}, Metadynamics\cite{meta_pnas,wtm} and more recently Variationally Enhanced Sampling\cite{omar2}.
However, identifying the appropriate CVs can at times require a lengthy, if instructive, process\cite{gervasioCV}. 

The other set of methods can be classified under the generic name of tempering. The precursor of this approach is called Parallel Tempering\cite{ptempering} (PT). In PT 
$M$ replicas of the same system at different temperatures are run in parallel. Periodically a Monte Carlo test is made and, if the test is successful, configurations are exchanged between replicas. 
The rationale for this approach is that at high temperature it is easier for the system to move from basin to basin and this information is carried down to the colder temperature by 
the Monte Carlo exchange process. This idea has been generalized and several tempering schemes have been proposed\cite{re1,re2,re3,wte,re4,bussi_cvt,re6,re7,re8,re9,re10} 
in which the Hamiltonians in each replicas are progressively modified\cite{ptempering} in order to favor the process of barrier crossing. Statistics is then collected in the unmodified replica.

Since it is relevant for what follows, we highlight the recent proposal made by Voth and collaborators\cite{Voth_sampling} and by our group\cite{quantumruge,dbms} 
to use artificially induced quantum effects to favor sampling. In this approach quantum effects are described by the isomorphism that is generated by the use of the 
Path Integral representation of Quantum Statistical Mechanics\cite{feynmanhibbs}. In this popular isomorphism each particle is mapped onto a polymer of $P$ beads interacting between themselves through a harmonic 
potential while the interaction between beads belonging to different polymers is appropriately reduced\cite{fcenter}.
 
The success of these attempts have stimulated us to reduce this approach to its bare essentials and somehow to generalize it. 
In order to reduce its computational cost the number of beads will be reduced to two, thus we shall abandon any pretense of describing each replicas as a representation, albeit approximate, 
of a quantum state. Having given ourselves such a freedom we shall also change the interaction between the beads. Thus our tempering scheme will be composed of $M$ replicas; in each replica every particle becomes a dimer, 
and the interaction that holds together the dimers is now anharmonic. By tightening the potential that holds the dimers together we can progressively come close to the Boltzmann distribution of the system. 
As in the Feynman's isomorphism, this is reached when the intra-dimer interaction becomes infinitely strong. Like in Ref. \onlinecite{dbms} a key to the success of this method is the use of Metadynamics to 
increase the dimer length fluctuations. Thus we shall refer to this method as Dimer Metadynamics (DM).

\section{Method}
As discussed in the Introduction, our method is inspired by previous attempts to use quantum effects to overcome 
sampling bottlenecks\cite{quantumruge,dbms}. However here we prefer to derive our formulas without 
making explicit reference to the Path Integral representation of Quantum Statistical Mechanics\cite{feynmanhibbs}.
We want to sample a Boltzmann distribution whose partition function is written as:
\begin{eqnarray}\label{z1}
Z_0 = \int dR \: \mbox{e}^{-\beta V(R)}
\end{eqnarray}
where $N$ particles of coordinates $R=\vec{r}_i, \: i=1...N$ interact via the $V(R)$ potential at the inverse temperature $\beta$.
Let us now introduce $3N$ new coordinates $X = \lbrace \vec{x}_i \rbrace_{i=1}^{N}$ and a function $F_\sigma (X)$ that depends parametrically on $\sigma$ and it is such that $ \int dX \mbox{e}^{-\beta F_\sigma (X)}=1$. 
We can now rewrite $Z_0$ as 
\begin{eqnarray}\label{zrew}
Z_0 = \int dRdX \: \mbox{e}^{-\beta F_\sigma (X)}\mbox{e}^{-\beta V(R)}
\end{eqnarray}
and making the coordinate transformation $X = R_1 - R_2$ and $ R = (R_1 + R_2)/2$ we obtain 
\begin{eqnarray}\label{zrew2}
Z_0 = \int dR_1dR_2 \: \mbox{e}^{-\beta F_\sigma (R_1 - R_2)}\mbox{e}^{-\beta V(\frac{R_1 + R_2}{2})} \:.
\end{eqnarray}
Of course the partition function in Eq. \eqref{zrew2} is fully equivalent to the Boltzmann distribution and the doubling of the coordinates is only apparent.

We now choose a family of functions $F_\sigma$ such that 
\begin{eqnarray}\label{dfam}
\lim_{\sigma\rightarrow 0} \mbox{e}^{-\beta F_\sigma (R_1 - R_2)} = \delta (R_1 - R_2)
\end{eqnarray}
where $\delta(R_1-R_2)$ is the Dirac's delta. We shall postpone the choice of $F_\sigma$ to later, in the meantime we 
note that if Eq. \eqref{dfam} holds for $\sigma \rightarrow 0$, we can make the approximation $V(\frac{R_1+R_2}{2}) \simeq \frac{V(R_1)}{2}+\frac{V(R_2)}{2}$ and 
rewrite the partition function as:
\begin{eqnarray}\label{zrew3}
Z_\sigma = \int dR_1dR_2 \: \mbox{e}^{-\frac{1}{2}V(R_1)}\mbox{e}^{-\beta F_\sigma(R_1-R_2)}\mbox{e}^{-\frac{1}{2}V(R_2)}
\end{eqnarray}
that is fully equivalent to the Boltzmann distribution (Eq. \eqref{z1}) in the limit $\sigma \rightarrow 0$, however, contrary to Eq. \eqref{zrew2} here the degrees 
of freedom are doubled in earnest. This partition function describes a system of dimers bound by $F_\sigma (R_1 - R_2)$ and interacting via the reduced potential $\frac{V(R)}{2}$.
At large values of $\sigma$ the dimers are loosely coupled and the two images of the system $R_1$ and $R_2$ can explore rather different configurations. By reducing $\sigma$ the coupling becomes 
stronger and stronger until for $\sigma\rightarrow 0$ the Boltzmann limit is approached.

We now devise a set of replicas of which the first one is $Z_0$ whose $\sigma$ we denote by $\sigma_0$ and the others are of the type in Eq. \eqref{zrew3} with progressively large values of $\sigma$, 
starting with $\sigma=\sigma_1$.
For reasons that will become apparent later we take $\sigma_0 = \sigma_1$.
The idea is then to set up a replica exchange scheme in which systems of different $\sigma_i, \: i=1,...,M$ are run in parallel and periodically Monte Carlo 
attempts at swapping the configurations are made\cite{Voth_sampling,dbms}.
In the tempering scheme proposed here the Monte Carlo test between neighboring replicas reads:
\begin{eqnarray}
p_{i,i+1}= \min \left[ 1, \mbox{e}^{-\beta\Delta E } \right] \label{pswp}
\end{eqnarray}
with:
\begin{flalign}
& \Delta E = \left[F_{\sigma_i}(R_1^{i+1} - R_2^{i+1})+F_{\sigma_{i+1}}(R_1^{i} - R_2^{i})\right] - \left[F_{\sigma_i}(R_1^{i} - R_2^{i})+F_{\sigma_{i+1}}(R_1^{i+1} - R_2^{i+1})\right] 
\end{flalign}
where the subscript in the many-body coordinates is the bead index of the dimer and the superscript identifies the replica index.
The Monte Carlo test between $Z_0$ and $Z_{\sigma_1}$ is given by:
\begin{flalign}
p_{0,1}=\min\left[1,\mbox{e}^{-\beta\Delta V^{(0,1)}}\right] \label{pacccg}
\end{flalign}
with:
\begin{flalign}
\Delta V^{(0,1)} = \left[ \frac{V(R_1^0)+V(R_2^0)}{2} + V\left(\frac{R_1^1+R_2^1}{2}\right)\right] - \left[ \frac{V(R_1^1)+V(R_2^1)}{2} + V\left(\frac{R_1^0+R_2^0}{2}\right)\right]
\end{flalign}
and we use the fact that we have chosen $\sigma_0 = \sigma_1$. 

We now turn to the choice of $F_\sigma (R_1 - R_2)$. We use the form 
\begin{eqnarray}
F_\sigma (R_1 - R_2 ) = \sum_{i=1}^N f_\sigma (\vec{r}_i^{\,1} - \vec{r}_i^{\,2})
\end{eqnarray}
where $\vec{r}_i^{\,1}$ and $\vec{r}_i^{\,2}$ are the coordinates of atom $i$ that has been split into the two beads $1$ and $2$ (see Eq. \eqref{zrew2}). For our approach to work  
it is necessary that for $\sigma \rightarrow 0$ the behavior of $\mbox{e}^{-f_\sigma (r)}$ is $\delta$-function like. One such class of functions can be obtained by considering for 
$0 < q \le 1$ the following representations of the three dimensional delta functions

\begin{eqnarray}
\delta (r) = \lim_{\sigma\rightarrow 0} \frac{\mbox{e}^{-\left[\left(1+\frac{r^2}{2q\sigma^2}\right)^q -1\right]}}{Z_{\sigma}^q}
\end{eqnarray}
where the normalization constant $Z_\sigma^q$ is given by
\begin{eqnarray}
Z_\sigma^q = 4\pi \int dr\: r^2 \mbox{e}^{-\left[\left(1+\frac{r^2}{2q\sigma^2}\right)^q -1\right]}
\end{eqnarray}
Using Eq. \eqref{dfam} and neglecting the immaterial constant $\log Z_\sigma^q$ we get for $f_\sigma^q(r)$
\begin{eqnarray}
f_\sigma^q (r) = \left(1+\frac{r^2}{2q\sigma^2}\right)^q -1
\end{eqnarray}
The choice of $f_\sigma^q (r)$ determines the potential with which two beads interact. For $q=1$ one has 
\begin{eqnarray}\label{fsgauss}
f_\sigma^{(1)} (\vec{r}\,) = \frac{r^2}{2\sigma^2} 
\end{eqnarray}
as in the standard Path Integral isomorphism. Otherwise for $0 < q < 1$ all $f_\sigma^q (r)$ exhibit a quadratic behavior 
$\frac{r^2}{2\sigma^2}$ at small $r$ and a slower growth at  larger distances. In particular, for $q=\frac{1}{2}$ $f_\sigma^q (r)$ grows linearly with $r$.
The transition between small-$r$ and large-$r$ regimes is controlled by the parameter $\sigma$, that in the spirit of the present work 
has no physical meaning and is only a tempering parameter.
As $q \rightarrow 0$ the asymptotic behavior is even slower and in the limit becomes logarithmic. However in this case $Z_\sigma^q \rightarrow \infty$ and the system becomes unstable.

In this first application of the method we choose $q=\frac{1}{2}$. This value is possibly not optimal but it is better than $q=1$. 
In our experimentation we found that $q=0.4$ and $q=0.3$ are also viable options. However the advantages did not seem so great as to warrant abandoning the more aesthetically pleasing $q=\frac{1}{2}$ choice.
In this respect we find amusing to note that also the quark-quark interaction has a linear asymptotic behavior.

The final and essential ingredient of our approach is the use of Metadynamics. Following Ref. \onlinecite{quantumruge} we shall combine the replica exchange scheme described above with 
Well-Tempered Metadynamics. In particular we shall use as CV the elastic energy  per particle stored in the dimer
\begin{eqnarray}\label{scv}
 s=\frac{1}{N \beta}\sum_{i=1}^{N}\left[\left(1+\frac{\left(\vec{r}_i^{\,2}-\vec{r}_i^{\,1}\right)^2}{2q\sigma^2}\right)^q-1\right]
\end{eqnarray}
The role of Metadynamics is to enhance the fluctuations of $s$ since in a Well-Tempered Metadynamics that uses $\gamma$ as boosting parameter the probability distribution 
of the biased variable, $p_b(s)$ is related to that in the unbiased ensemble by the relation $p_b(s) \propto \left[p(s)\right]^{1/\gamma}$.
In our case since $s$ is related to the elastic energy, configurations in which the dimer is highly stretched are more likely to be observed.

\section{Results}
\begin{figure}[h]
\includegraphics*[width=5.6in]{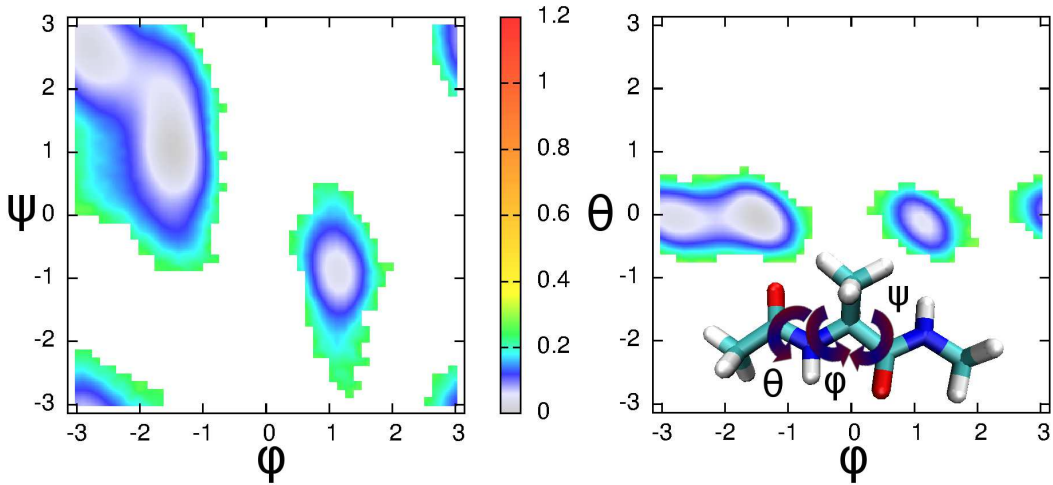}
\caption{Free energy surfaces in eV of Alanine Dipeptide as function of the dihedral angles defined in the right panel. The temperature was 300 K, 
the six replicas had $\sigma_i=$0.02, 0.02, 0.06, 0.15, 0.6 and 3.0 \AA. The simulation was 100 ns long with a timestep of $dt=1$ fs and swaps between configurations were attempted every 6 ps.
Each 2 ps a Metadynamics Gaussian was deposited with initial height $w_0 =300$ K and bias factor $\gamma = 7$, the width of the Gaussians depended on the replica index and 
were $\sigma_g=$ 9.8, 9.8, 5.1, 4.7, 4.3 and 3.9 meV}\label{ala2pic}
\end{figure}

Before discussing the two applications presented here we want briefly to substantiate the assertion made earlier that the choice $q=\frac{1}{2}$ 
is more efficient than $q=1$. For this reason we consider once more the case of Alanine Dipeptide, a classical simple example on which 
new sampling methods are often tested. In \ref{ala2pic} we show the results obtained with the DM method with $q=\frac{1}{2}$ where only 6 replicas 
where required while in contrast, for $q=1$ we had to use 10 replicas.

\begin{figure}[h]
\includegraphics*[width=5.6in]{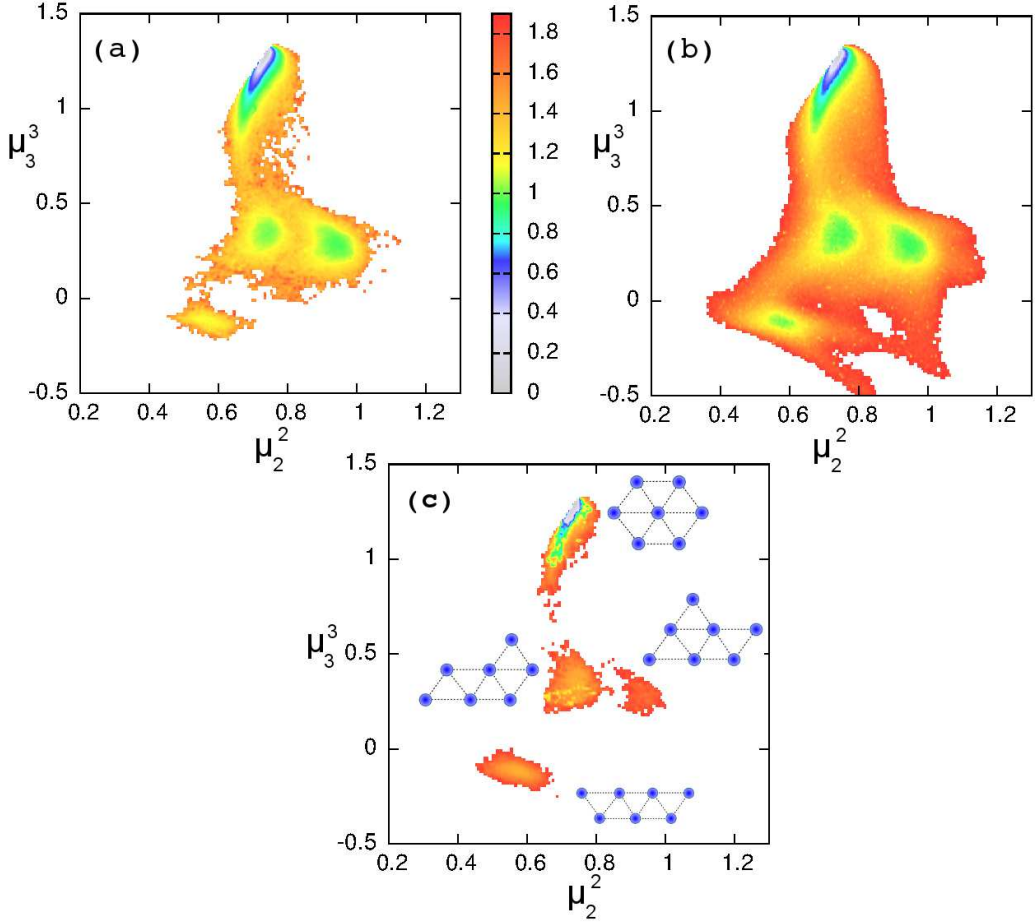}
\caption{Free energy surface of a two dimensional LJ cluster of 7 atoms as function of the second and third momentum of the coordination number in Lennard Jones units obtained with 
(a) DM and (b) standard Metadynamics at a temperature $T=0.1 \epsilon$.
In (c) the result obtained with DM at $\sigma_1=0.01$ without using replica exchange.}\label{figljfes}
\end{figure}

We now tackle a simple, yet challenging two dimensional system composed of 7 atoms interacting via Lennard-Jones (LJ) potential. 
This cluster is known to have three metastable configurations that can be represented as local minima in the free energy expressed as a function of  
the second and third momentum of the coordination number\cite{tribello} as shown in \ref{figljfes}.
Standard MD is not able to sample the metastable states of this simple cluster in practical times and enhanced sampling methods have been used to study it\cite{tribello}. In \ref{figljfes} 
the free energy surface of this LJ cluster at $T = 0.1 $ has been computed with DM using 5 replicas with $\sigma_i = $0.01, 0.01, 0.05, 0.2 and 0.6, where $\sigma_0 = \sigma_1$ 
as discussed before and LJ units are used. The timestep was $dt = 0.01$ and the simulation ran for $10^8$ steps. 
Swaps between replicas were tried every 500 steps and the Metadynamics Gaussians were deposited every 
 200 steps with initial height of $0.1$ and width depending on the replica index, $\sigma_g =$ 5.7, 3.6, 2.1  and 1.4; the bias factor was $\gamma = 2.71$. 
 
\begin{figure}[h]
\includegraphics*[width=5.6in]{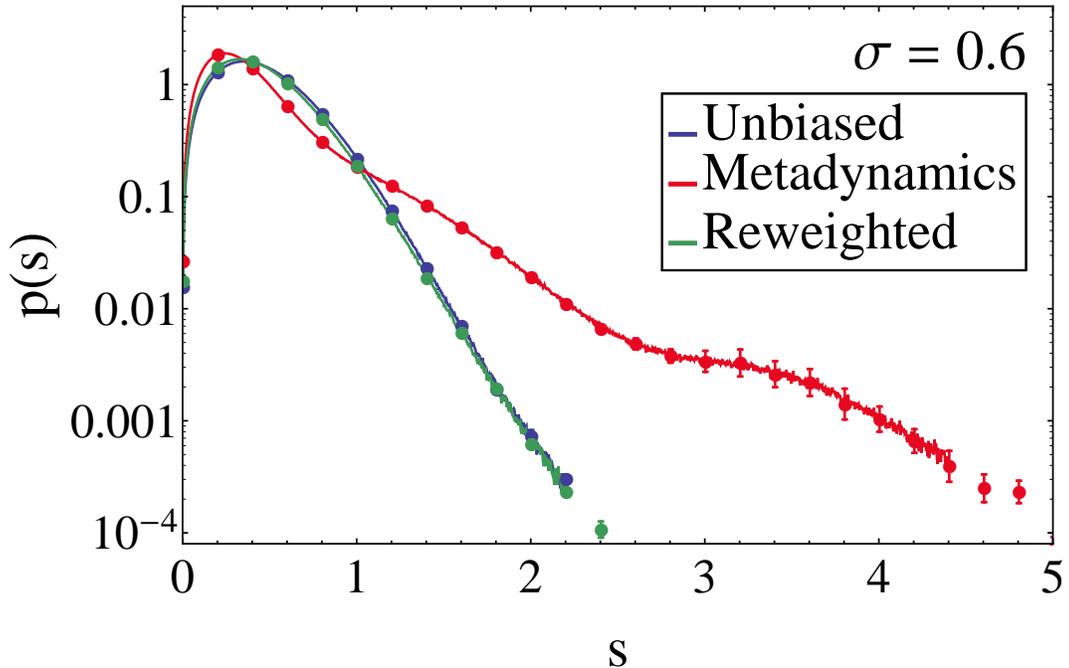}
\caption{Probability distribution of the length of the dimer for a two 
dimensional Lennard-Jones cluster of 7 atoms obtained from an unbiased simulation (blue line), with Metadynamics on 
the spring energy (red line) and the unbiased distribution recovered from the biased simulation (green line). 
}\label{geom}
\end{figure}

 These results are compared to a $10^8$ steps long Metadynamics\cite{wtm} simulation with timestep $dt = 0.005$ and bias factor $\gamma = 10$, where every $500$ steps 
 the second and third momentum of the coordination number are biased with Gaussians of initial height $w_0= 0.01$, width $\sigma = (0.02 ; 0.02)$.
 As shown in \ref{geom}
DM dramatically enhances the sampling of the long distances tail of the probability distribution of the dimer length. Analogously to the Path Integral case\cite{quantumruge,dbms}, this effect 
increases the delocalization of the particle and indeed even without replica exchange DM can locate all of the four minima in the free energy surface (\ref{figljfes}(c)). 

We consider now the more complex case of Alanine Tripeptide in vacuum as described by the Charmm22$^\star$ \cite{charmm22s} forcefield, a protein that can have different conformations 
separated by moderately high energy barriers.
%In \ref{ala1d} we compare two free energy surfaces of Alanine Tripeptide, each respectively as function of the dihedral angle $\phi_1$ and $\phi_2$ defined in the same picture. The 
%results obtained with the Dimer method are equal within statistical error to those obtained with Parallel Tempering\cite{ptempering} (PT) and with Variationally Enhanced Sampling\cite{omar2} (VES). 
\begin{figure}[h]
\includegraphics*[width=5.6in]{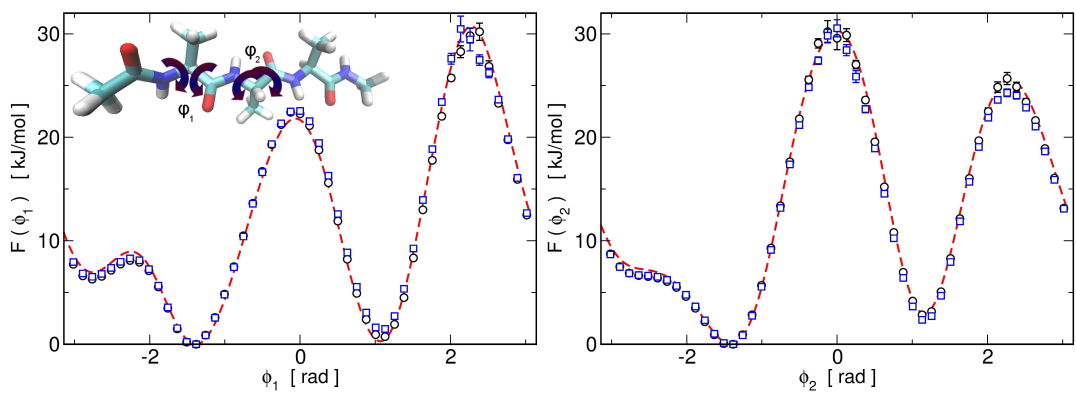}
\caption{Free energy as function of the dihedral angles $\phi_1$ (left) and $\phi_2$ (right). Black circles are results from the $\sigma=0.01$~\AA~  DM replica, blue squares are obtained from the Boltzmann DM replica and 
dashed line from VES.
}\label{ala1d}
\end{figure}
These results were obtained with DM using 4 replicas with interaction strengths $\sigma =$ 0.01, 0.05, 0.08, and 0.15 \AA~ plus one additional Boltzmann replica (Eq. \eqref{zrew2}) with $\sigma_0=\sigma_1$. The temperature was $T = 300$ K and the simulation 
ran for 100 ns with a timestep of $dt = 2$ fs. Each 2 ps a Gaussian was deposited with initial height $w_0 = 300$ K and bias factor $\gamma = 4.3$, the standard deviations of the 
Gaussians depended on the replica index and were $\sigma_g = $ 33.5, 22.3, 13.4, 6.7 and 4.5 meV. Swaps between replicas were attempted every 5 ps.
\begin{figure}[h]
\includegraphics*[width=5.6in]{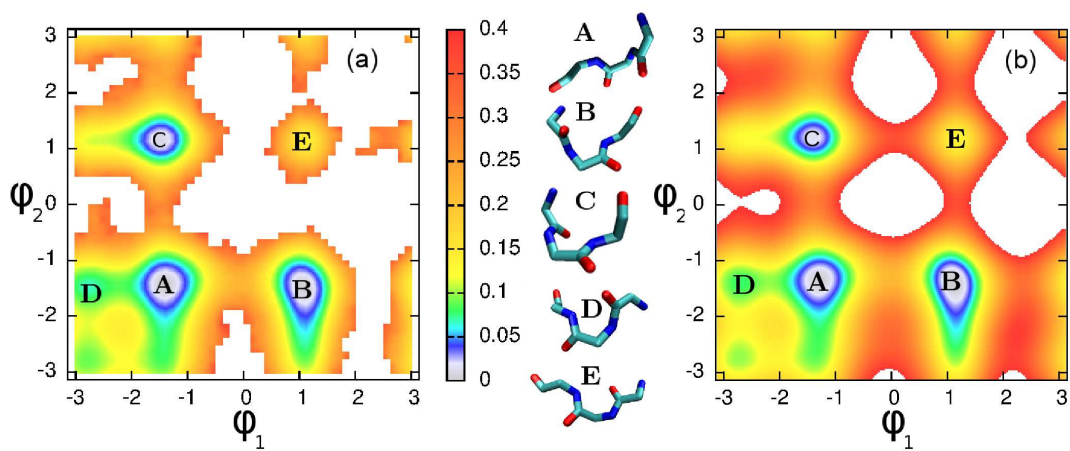}
\caption{Free energy surface in eV of Alanine Tripeptide obtained with (a) DM for the $\sigma_0 = 0.01$~\AA~ replica and (b) VES. For each of the minima a representative sample of the 
sampled configurations is shown.
}\label{ala2d}
\end{figure}
%The PT simulation was unbiased and ran for 1 $\mu$s with a timestep $dt = 1$ fs, it used 6 replicas at temperatures $T_i =$ 300, 346, 420, 504, 610 and 750 K and attempted to swap configurations every 5 ps. 

We calculate a reference free energy surface by using the Variationally Enhanced Sampling (VES) of Ref. \onlinecite{omar2} 
in the Well-Tempered variant of Ref. \onlinecite{Valsson-JCTC-2015}. We employed as CVs the three dihedral 
angles $\Phi_{1},\Phi_{2},\Phi_{3}$ and expanded the bias potential in a Fourier series 
of size 6 for each CV, resulting a total number of 2196 basis functions. To optimize the VES functional we 
used the method of Bach\cite{Bach-NIPS-2013}. The coefficients were updated every 1 ps and a 
fixed step size of 0.08 kJ/mol was used. In order to achieve a Well-Tempered distribution we use a 
bias factor of 10 and the target distribution was updated every 500 ps\cite{Valsson-JCTC-2015}. 
The simulation cost was equivalent to that of a 50 ns Molecular Dynamics run.

The results of these two calculations are compared in \ref{ala1d} in which the free energy as function of the dihedral angle $\phi_1$ 
and of $\phi_2$ are shown along with the definition of the angles. The agreement between the two calculations is excellent and we note that strictly speaking 
use of the rigorously Boltzmannian replica $Z_0$ is not necessary. Also the correlations between $\phi_1$ and $\phi_2$ (\ref{ala2d}) are well represented as well as the location and population of the different conformers.
We underline the fact that in DM as opposed to VES no CVs need to be introduced.

\section{Conclusions}
Taking inspiration from de Broglie swapping Metadynamics\cite{dbms} and from Ref. \onlinecite{Voth_sampling}, we have used artificial delocalization effects to enhance sampling of Boltzmann systems. 
The delocalization has been obtained by mapping each particle into a dimer in which atoms are bound by an anharmonic potential. 

The computational cost relative to previous simulation methods is reduced. In fact here we deal only with dimers and not with polymers as in Ref. \onlinecite{dbms} and also 
the more gentle behavior of $f_\sigma^{\frac{1}{2}}(r)$ at large distances favors conformational swaps reducing the number of replicas needed.

Like previous Path-Integral-based methods\cite{dbms,quantumruge} and PT, DM does not require choosing a CVs. Furthermore it offers 
a natural way of enhancing sampling of only a part of the system. For instance the conformational landscape of a mobile loop in a large protein could be selectively targeted.

%\begin{figure}[b]
%\includegraphics*[width=5.6in]{images/toc.eps}
%\caption{For table of contents only.
%}
%\end{figure}

\begin{acknowledgement}
All calculations were performed on the Brutus HPC cluster at ETH Zurich and on the Piz Dora supercomputer at the Swiss National Supercomputing Center (CSCS) under project ID u1. 
We acknowledge the European Union Grant ERC-2014-Adg-670227 and Marvel 51NF40\_141828. 

We would also like to acknowledge Omar Valsson for his help with VES on Alanine Tripeptide.
\end{acknowledgement}

\bibliography{pimetad}
\end{document}